\documentclass[runningheads,a4paper]{llncs}

\usepackage{amssymb}
\usepackage{graphicx}
\usepackage{times}
\usepackage{footmisc}
\usepackage[ruled,vlined,linesnumbered]{algorithm2e}

\usepackage{pstricks}
\usepackage{pst-node}
\usepackage{pst-tree}
\usepackage{longtable}

\usepackage{url}
\urldef{\mailsa}\path|{alfred.hofmann, ursula.barth, ingrid.haas, frank.holzwarth,|
\urldef{\mailsb}\path|anna.kramer, leonie.kunz, christine.reiss, nicole.sator,|
\urldef{\mailsc}\path|erika.siebert-cole, peter.strasser, lncs}@springer.com|    
\newcommand{\keywords}[1]{\par\addvspace\baselineskip
\noindent\keywordname\enspace\ignorespaces#1}

\begin{document}

\mainmatter  

\title{Interpretation of  \textit{NDTM} in the definition of \textit{NP}}

\titlerunning{Interpretation of  \textit{NDTM} in the definition of \textit{NP}}

%
%
\author{JianMing ZHOU, Yu LI\inst{1,2}}
\authorrunning{JianMing ZHOU, Yu LI}

\institute{MIS, Universit\'e de Picardie Jules Verne, Amiens, France, yu.li@u-picardie.fr \and Institut of computational theory and application, Huazhong University of Science and Technology, Wuhan, China
}
%
%

\toctitle{Lecture Notes in Computer Science}
\tocauthor{Authors' Instructions}
\maketitle

\begin{abstract}

In this paper, we interpret  \textit{NDTM} (NonDeterministic Turing Machine) used to define \textit{NP} by tracing to the source of   \textit{NP}. Originally \textit{NP} was defined as the class of problems solvable in polynomial time  by a \textit{NDTM} in Cook's theorem,  where the \textit{NDTM} was represented as \textit{Query Machine} of essence \textit{Oracle}. Later a model  consisting of a guessing module and a checking module was proposed to replace the \textit{NDTM}.  This model of essence  \textit{TM} has a fundamental difference from the \textit{NDTM}   of essence \textit{Oracle},   but people  still use the term  \textit{NDTM} to designate this model, which leads to the disguised displacement of   \textit{NDTM}  and  produces out the  verifier-based definition of  \textit{NP}   as the class of problems verifiable in polynomial time  by a \textit{TM} (Turing Machine). This  verifier-based one   has been  then accepted as the standard definition of  \textit{NP} where comes from   the famous equivalence of the two definitions of  \textit{NP}. Since then the notion of \textit{nondeterminism}   is lost from \textit{NP}, which causes ambiguities in understanding  \textit{NP}  and then great difficulties in  solving  the \textit{P versus NP}  problem. 

Since \textit{NP} is originally related with \textit{Oracle} that comes from Turing's work about \textit{Computability}, it seems quite necessary to trace back to Turing's work and clarify further the issue about \textit{NP}.

\end{abstract}

\section{Introduction}

The \textit{P versus NP}  problem was selected as one of the seven millennial challenges by the Clay Mathematics Institute  in 2000 \cite{cook2}. This problem goes far beyond the field of computer theory  and penetrates  into mathematics, mathematical logic, artificial intelligence, and even becomes the basic problem in philosophy. In introducing the second  poll about  \textit{P versus NP} conducted by Gasarch in 2012 \cite{william}, Hemaspaandra said: \textit{I hope that people in the distant future will look at these four articles to help get a sense of peopleÕs thoughts back in the dark ages when P versus NP had not yet been resolved.} 

$P$ stands for \textit{Polynomial time}, meaning that a problem in $P$ is solvable by a deterministic Turing machine in polynomial time. Concerning the definition of $NP$, the situation is much more complex,   \textit{NP} stands for \textit{Nondeterminisitc Polynomial time}, meaning that  a problem in \textit{NP} is  solvable  by a Nondeterministic Turing machine in Polynomial time \cite{cook1}\cite{cook2}. However, this solver-based definition is considered academically as equivalent with another verifier-based definition of  \textit{NP}   \cite{np}:

\textit{The two definitions of NP as the class of problems solvable by a nondeterministic Turing machine in polynomial time and the class of problems verifiable by a deterministic Turing machine in polynomial time are equivalent. The proof is described by many textbooks, for example Sipser's Introduction to the Theory of Computation, section 7.3.}\\

Due to this equivalence, the verifier-based definition  has been accepted as the standard definition of  \textit{NP},   the  \textit{P versus NP} problem is then  stated as: 
\begin{itemize}
\item    $P \subseteq NP$, since a  problem solvable by a \textit{TM}  in polynomial time  is verifiable by a \textit{TM}  in polynomial time. 
\item  $NP = P$? whether a problem verifiable by a \textit{TM}  in polynomial time is   solvable by a a \textit{TM}  in polynomial time?  
\end{itemize}

 In this paper, by tracing the source of  \textit{NP}, we investigate \textit{NDTM}  used to define \textit{NP}  and reveal the disguised displacement  of   \textit{NDTM}, which produces out the verifier-based definition of  \textit{NP} as well as the equivalence of the two definitions of \textit{NP}.

The paper is organized as follows: we return to the origin of \textit{NDTM}  in Section 2,  examine its change in Section 3, analyze the proof of   the equivalence of   the two definitions of \textit{NP}  in Section 4, and conclude the paper in Section 5.
 
\section{ \textit{NDTM} as  \textit{Oracle}}

\textit{NDTM}  was  formally used to define \textit{NP} in  Cook's paper entitled   \textit{The complexity of theorem proving procedures}   \cite{cook1}.

 \subsection{ \textit{NDTM} in  Cook's theorem}  
 
 Cook's theorem was originally stated as  \cite{cook1}:

\textbf{Theorem 1}
\textit{If a set $S$ of strings is accepted by some nondeterministic Turing machine within polynomial time, then $S$ is $P$-reducible to \{DNF tautologies\}}.  \\

Here $S$ refers to a set of  instances   of a  problem that have solutions,  which later becomes the solver-based definition of $NP$ in terms of   \textit{language}  \cite{cook2}: 

\textit{A problem in NP is a language accepted by some nondeterministic Turing machine within polynomial time}. 

Concerning \textit{\{DNF   tautologies   $ \neg A(w)$\}},    it  can be transformed into  \textit{\{CNF  satisfiabilities    A(w)\}}, so it corresponds to  the \textit{SAT}  problem. \\ 

 \textbf{Theorem 1} is nowadays expressed as  \cite{np}:
 
 \textbf{Cook's theorem}
\textit{A problem in  NP   can be reduced  to the SAT  problem by a (deterministic) Turing machine  in polynomial time.}

 \subsection{Analysis of  \textit{Query Machine} }  
 
 The main idea of the proof of  \textbf{Theorem 1}  is to  construct  $A(w)$ to express that  a set $S$ of strings is accepted by a \textit{NDTM}   in polynomial time \cite{cook1}:

 \textit{Suppose a nondeterministic Turing machine $M$ accepts a set $S$ of strings within time $Q(n)$, where $Q(n)$ is a polynomial. Given an input $w$ for $M$, we will construct a propositional formula $A(w)$ in conjunctive normal form ($CNF$) such that $A(w)$ is satisfiable iff $M$ accepts $w$.  Thus $\neg A(w)$ is easily put in disjunctive normal form (using De MorganÕs laws), and $\neg A(w)$ is a tautology if and only if w $ \not \in S$. Since the whole construction can be carried out in time bounded by a polynomial in $\mid w \mid$ (the length of $w$), the theorem will be proved.}\\
 
This  \textit{NDTM}   is then represented as \textit{Query Machine} \cite{cook1}:
 
  \textit{By reduced we mean, roughly speaking, that if tautology hood could be decided instantly (by an "oracle") then these problems could be decided in polynomial time. In order to make this notion precise, we introduce query machines, which are like Turing machines with oracles in [1].} \\
  
 This  \textit{query machine} is described as \cite{cook1}:

  \textit{A query machine is a multitape Turing machine with a distinguished tape called the query tape, and three distinguished states called the $query~state$, $yes~state$, and $no~state$, respectively. If $M$ is a query machine and $T$ is a set of strings, then a $T$-computation of $M$ is a computation of $M$ in which initially $M$ is in the initial state and has an input string $w$ on its input tape,  and each time $M$ assures the query state there is a string $u$ on the query tape,  and the next state $M$ assumes is the yes state if $u \in  T$ and the no state if $u  \not  \in T$.  We think of an 'oracle', which knows $T$, placing $M$ in the yes state or no state}.  \\
  
   \begin{figure} [h]
\begin{center}
\includegraphics[scale=.4]{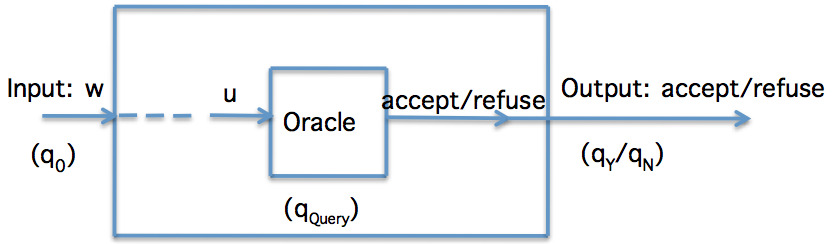}
\end{center}
\caption{A computation of   \textit{NDTM} as \textit{Oracle}}
\label{fig5}
\end{figure}

  The set of T of strings is explained as  \cite{cook1}:
  
  \textit{Definition. A set S of strings is P-reducible (P for polynomial) to a set T of strings iff there is some query machine M and a polynomial Q(n) such that for each input string  w, the T-computation of M with input w halts within Q($\mid w \mid$) steps ($\mid w \mid$ is the length of w) and ends in an accepting state iff  $w    \in S$}. 

\textit{It is not hard to see that P-reducibility is a transitive relation. Thus the relation $E$ on sets of strings, given by $(S,T)  \in E$ iff each of $S$ and T is P-reducible to the other, is an equivalence relation.  The equivalence class containing a set S will be denoted by deg (S) (the polynomial degree of difficulty of S).}\\

We  use  the  graph isomorphism problem cited in  \cite{cook1}  to help interpreting \textit{Query Machine}.

 \textbf{Example: Graph isomorphism problem}

Given two finite undirected graphs $G_1$ and $G_2$, the   problem consists in determining whether $G_1$ is isomorphic to $G_2$. 

An isomorphism of $G_1$ and $G_2$ is a bijection $f$ between the vertex sets of $G_1$ and $G_2$, $f : V(G_1) \rightarrow V(G_2)$, such that any two vertices $u$ and $v$  are adjacent in  $G_1$ if and only if $f(u)$ and $f(v)$ are adjacent in $G_2$. In this case, a solution to an instance refers to an isomorphism between $G_1$ and $G_2$.\\

We give  the following  two instances. \textit{Instance 1}: A pattern graph $G_{p1}=(V_{p1}, E_{p1})$ and a text graph $G_{t1}=(V_{t1}, E_{t1})$,  \textit{Instance 2}: A pattern graph $G_{p2}=(V_{p2}, E_{p2})$ and a text graph $G_{t2}=(V_{t2}, E_{t2})$.

\begin{figure} [h]
\begin{center}
\includegraphics[scale=.5]{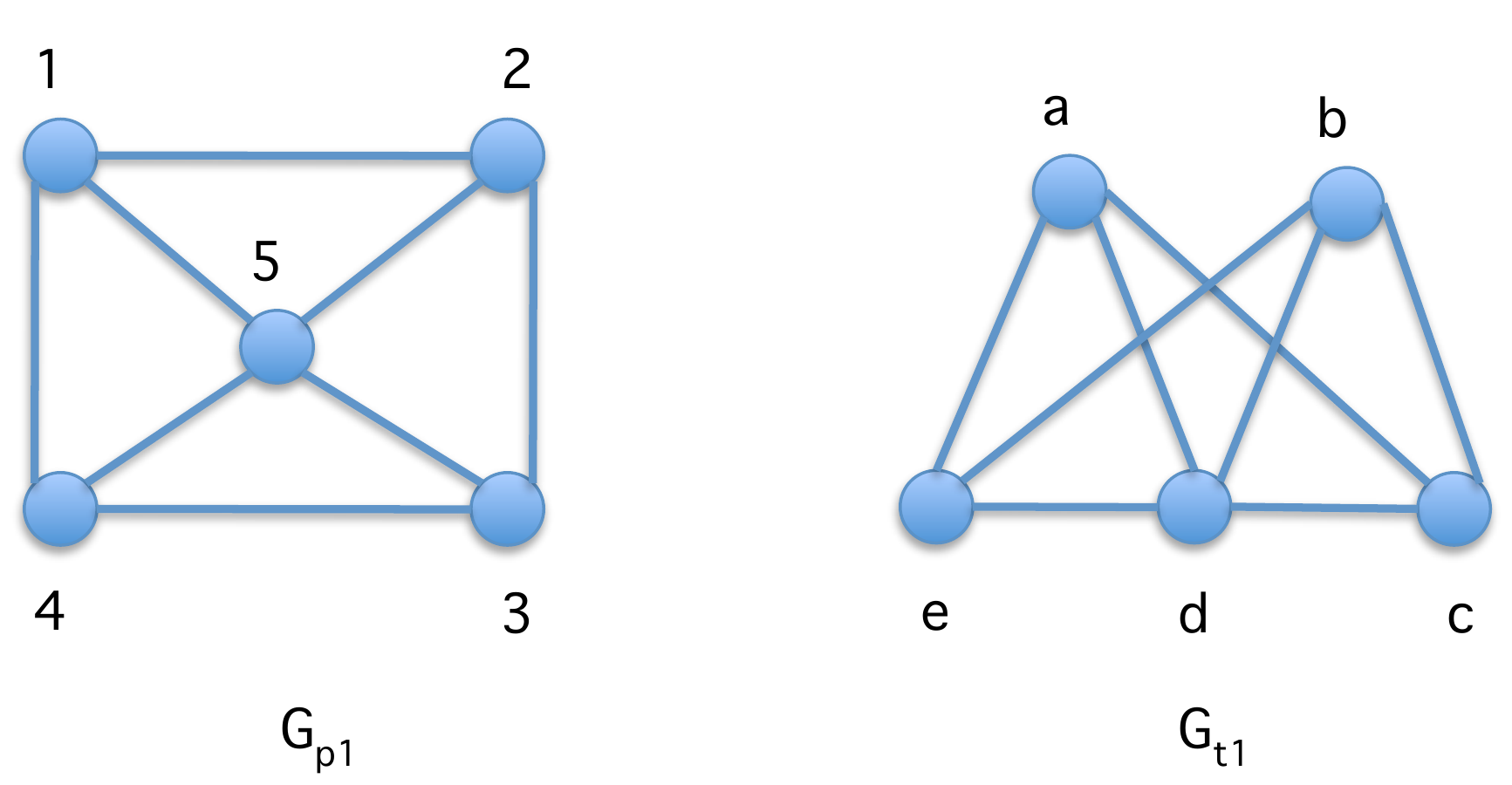}
\end{center}
\caption{Instance 1}
\label{fig1}
\end{figure}

\begin{figure} [h]
\begin{center}
\includegraphics[scale=.5]{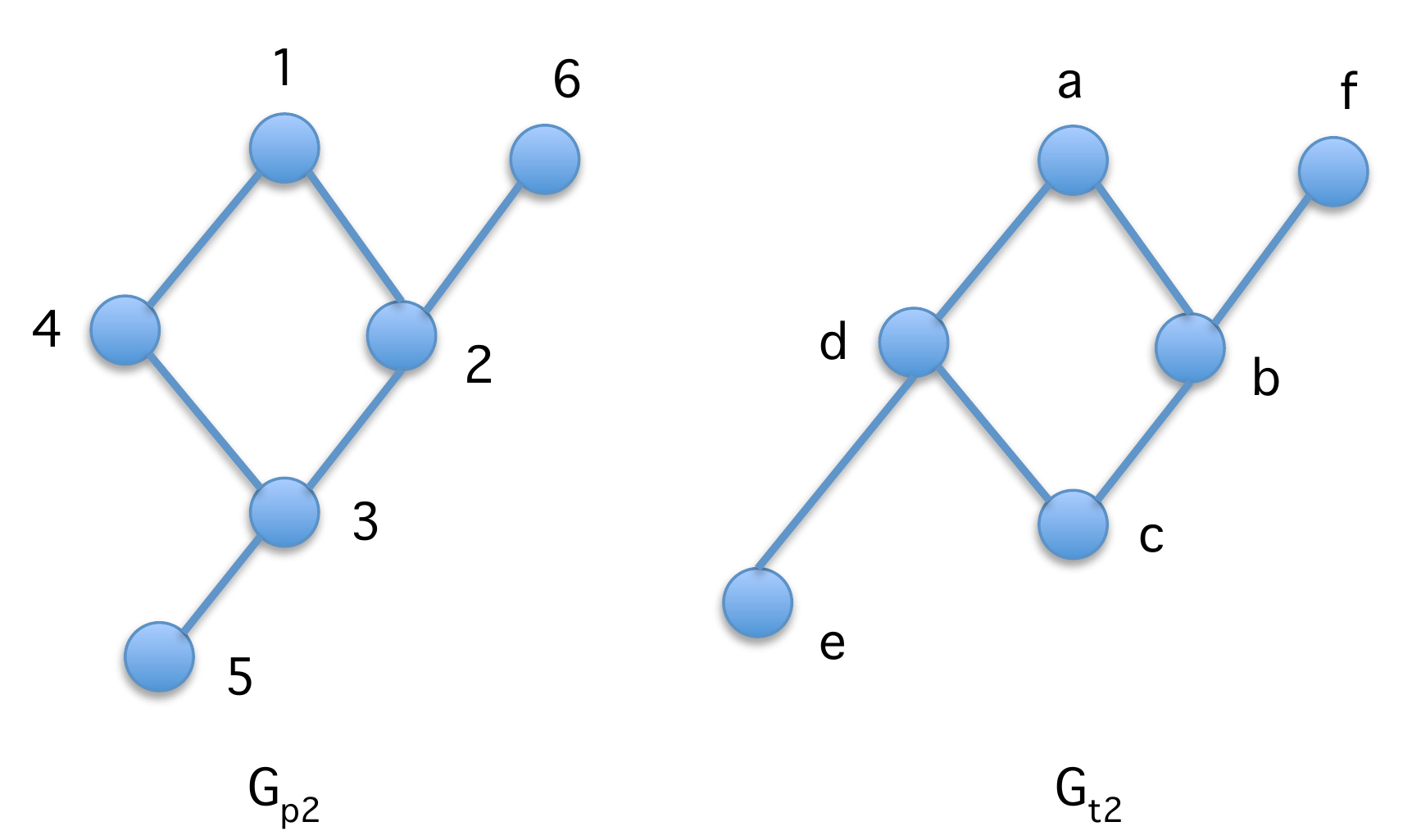}
\end{center}
\caption{Instance 2}
\label{fig2}
\end{figure}

For \textit{Instance 1},    $G_{p1}$ is  isomorphic to  $G_{t1}$, as there exists an isomorphism :   $f(1)=a, f(2)=b, f(3)=c, f(4)=e, f(5)=d$; while for  \textit{Instance 2},  $G_{p2}$ is not isomorphic to  $G_{t2}$, as there is not any isomorphism of $G_{p2}$ and  $G_{t2}$. \\

Let us analyze how a \textit{query machine}  $M$ accepts a set S of strings  in polynomial time (Fig.3). Initially, $M$ is in the initial state $q_0$ and has  $w$   as input representing an instance of a problem. Then, $M$   assures the query state $q_{Query}$ where there is a string $u$ representing $w$ by a formula in terms of $CNF$. $u$  is taken as input of an \textit{oracle} and this \textit{oracle} instantly determines whether $u \in  T$, that is, whether $u$ is  satisfiable. Finally, according to the  obtained reply,   if $u \in  T$ then  the \textit{oracle} places $M$ in the yes state $q_Y$ and accepts $w$;   or if $u  \not  \in T$ then   the \textit{oracle} places $M$ in the no state $q_N$ and refuses $w$.

For the graph isomorphism problem, $S$ refers to a set of strings that represents all instances that have solutions, for example,  $S = \{\overline {G_{p1}}\ast\ast \overline {G_{t1}},  \ldots\}$. Note that $S$ does not contain   $\overline {G_{p2}}\ast\ast \overline {G_{t2}}$, because Instance 2 has no solution. $T$ refers to the corresponding set  of $CNF$ formulas that are  satisfiable.  
$M$ accepts $w=\overline {G_{p1}}\ast\ast \overline {G_{t1}}$, but refuses $w=\overline {G_{p2}}\ast\ast \overline {G_{t2}}$. \\ 

Therefore, saying that a \textit{query machine} accepts a set $S$ of strings in polynomial time, in fact that is to say  that an \textit{oracle}  accepts a set $S$ of strings in polynomial time. 

In other words,  the essence of  the  \textit{NDTM}  in Cook's theorem  is  \textit{Oracle}.

\section{ \textit{NDTM}  as \textit{TM} }
 
However,   \textit{Oracle}  is only a concept in thought experiments borrowed  by Turing in his doctoral dissertation  with the intention to represent   something opposed to     Turing Machine (\textit{TM}) of  essence \textit{Computability}  \cite{martin}\cite{turing}, so it   cannot carry out any real computation.  Therefore, later researchers   proposed a  \textit{NDTM}  model  to replace the  \textit{NDTM} of essence  \textit{Oracle}.\\

In Garey and Johnson's \textit{Computers and Intractability}   \cite{garey}, this model is presented as:

\textit{The NDTM model we will be using has exactly the same structure as a DTM  (Deterministic Turing Machine), except that it is augmented with a guessing module having its own write-only head.}

A computation of such a machine   takes place in two distinct stages (see  \cite{garey}, p. 30-31):

\textit{The first stage is the "guessing" stage. Initially, the input string $x$ is written in tape squares 1 through $\mid x \mid $ (while all other squares are blank), the read-write head is scanning square 1, the the write-only head is scanning square -1, and the finite state control is "inactive". The guessing module then directs the write-only head, one step at a time, either to write some symbol from $\Gamma$ in the tape square being scanned and move one square to left, or to stop, at which point the guessing module becomes inactive and the finite state control is activated in state $q_0$. The choice of whether to remain active, and, if so, which symbol from $\Gamma$ to write, is made by the guessing module in a totally arbitrary manner. Thus the guessing module can write any string from $\Gamma*$ before it halts and, indeed, need never halt.}

\textit{The "checking" stage begins when the finite state control is activated in state $q_0$. From this point on, the computation proceeds solely under the direction of the NDTM program according to exactly the same rules as for a DTM. The guessing module and its write-only head are no longer involved, having fulfilled their role by writing the guessed string on the tape. Of course, the guessed string can (and usually will) be examined during the checking stage. The computation ceases when and if the finite state control enters one of the two halt states (either $q_Y$ or $q_N$) and is said to be an accepting computation if it halts in state $q_Y$. All other computations, halting or not, are classed together simply as non-accepting computations. } 

 \begin{figure} [h]
\begin{center}
\includegraphics[scale=.3]{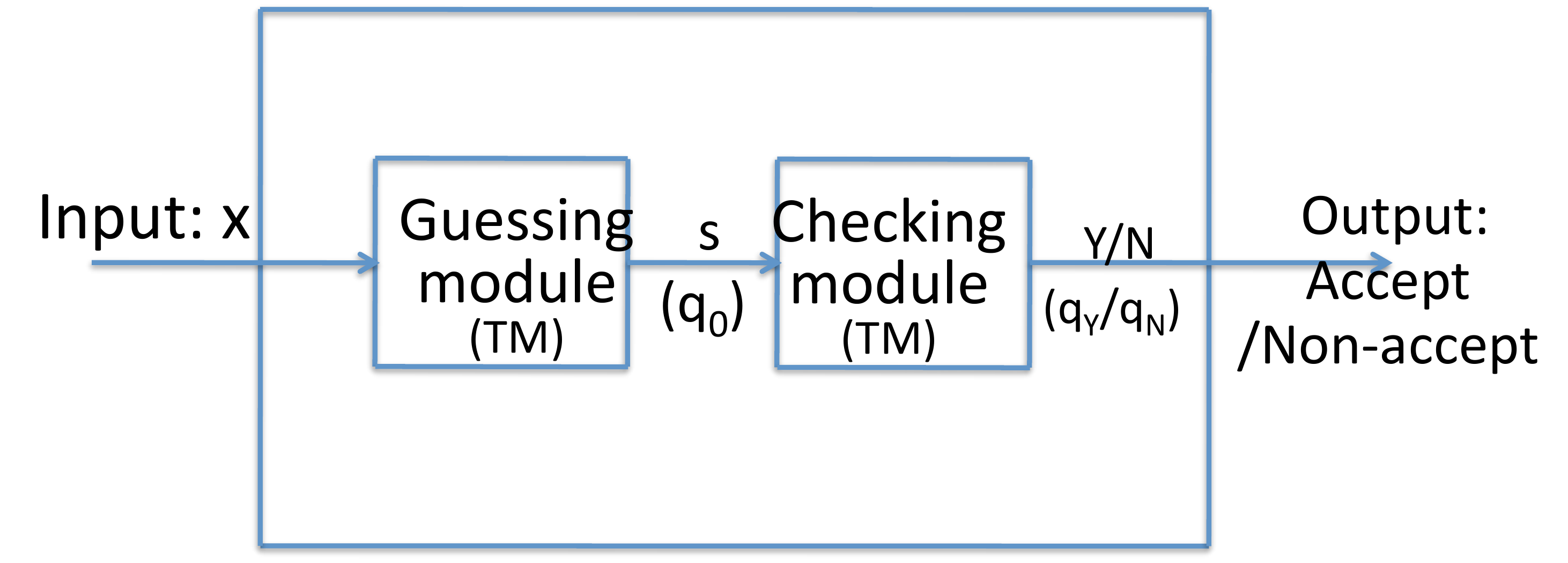}
\end{center}
\caption{A computation of the \textit{NDTM} as \textit{TM}}
\label{fig3}
\end{figure}

For a given instance $x$,  a guessing module finds a certificate $s$ of solution, then $s$ is  verified by a  checking module.  If  $s$ is a solution, the computation halts in state $q_Y$ and the machine can determine  that $x$ has a solution. However  if $s$ is not a solution,  the machine  can  determine neither that   $x$ has no  solution, nor  that  $x$ has a  solution.  In other words,  the  state $q_N$ in Fig.4 is \textit{nondeterministic}.

For \textit{Instance 2}, if a certificate  with $f(1)=a, f(2)=b, f(3)=c, f(4)=d, f(5)=e,  f(6)=f$ is generated by the guessing module,   and it is checked out to not be a solution, then  the machine can determine neither that \textit{Instance 2} has no solution, nor that \textit{Instance 2} has a solution. \\
 
 This \textit{NDTM}  is   actually  described   as \cite{sipser}: 

\textit{At any point in a computation the machine may proceed according to several possibilities. The computation of a nondeterministic Turing machine is a tree whose branches correspond to different possibilities for the machine. If some branch of the computation leads to the accept state, the machine accepts its input.}\\

The  essence of this \textit{NDTM} is   \textit{TM}, which  is  confirmed  in \cite{sipser}:

\textit{
\textbf{Theorem 3.16}
Every nondeterministic Turing machine has an equivalent deterministic Turing machine.\\
} 

Therefore,  this  \textit{NDTM}  of essence  \textit{TM} is completely different from   that \textit{NDTM}  of essence  \textit{Oracle} in Fig.3. Unfortunately,  people do not realize this  fundamental difference,  and still use the same term  \textit{NDTM}  to designate two different concepts. Consequently  \textit{TM}  is confused with  \textit{Oracle}, and it produces out  the following famous equivalence of the two definitions of \textit{NP}.

   \section{Analysis of   the equivalence  of  the two definitions of \textit{NP} }  
   
   Let us  analyze the proof described  in  Sipser's \textit{Introduction to the Theory of Computation}   (section 7.3) \cite{sipser}:
   
      \subsection{Description of the proof}

\textit{ \textbf{Theorem 7.20} 
A language is in $NP$ iff it is decided by some nondeterministic polynomial time Turing machine.}

\textit{
\textbf{Proof idea}:
We show how to convert a polynomial time verifier to an equivalent polynomial time NDTM and vice versa. The NDTM simulates the verifier by guessing the certificate. The verifier simulates the NDTM by using the accepting branch as  the certificate.
} 

\textit{
\textbf{Proof}:
From the forward direction of this theorem, let $A$ in \textit{NP} and show that $A$ is decided by a polynomial time $NDTM$ $N$. Let $V$ be the polynomial time verifier for $A$ that exists by the definition of \textit{NP}. Assume that $V $is a $TM$ that runs in time $n^k$ and construct $N$ as follows.
} 

\textit{
N = On input w of length n:
 \begin{enumerate}
\item Nondeterministically select string $c$ of length at most $n^k$.
\item Run $V$ on input $<w, c>$.
\item If $V$ accepts, accepts; otherwise, reject.
 \end{enumerate}    
To prove the other direction of the theorem, assume that $A$ is decided by a polynomial time $NDTM$ N and construct a polynomial time verifier $V$ as follows:
V = On input $<w,c>$, where $w$ and $c$ are strings:
 \begin{enumerate}
\item  Simulate $N$ on input $w$, treating each symbol of $c$ as a description of nondeterministic choice to make at each step.
\item If this branch of N's computation accepts, accept; otherwise, reject.
 \end{enumerate}
 }

     \subsection{Analysis of the proof }  

According to the proof idea,  the proof is based on the equivalence between the \textit{verification} of a certificate $c$ by   $V$ and the  \textit{decision}  for accepting instance $w$ by   $NDTM$ $N$:
\textit{
 \begin{itemize}
\item From the forward direction of the theorem: If V accepts, accepts; otherwise, reject;
\item From the other direction of the theorem: If this branch of N's computation accepts, accept; otherwise, reject.
 \end{itemize} 
  }

In fact this equivalence is   premised,  it holds only  with   \textit{NDTM} of essence  \textit{Oracle}  in Fig.3 where the verifier $V$ checks the result obtained by $Oracle$,  the \textit{verification} is  certainly consistent with the \textit{decision}, then the \textit{verification} and the \textit{decision} are  equivalent. However, the situation is completely different with the  \textit{NDTM}  in Fig.4, that is, in this proof.  \\

Let us look at the \textit{HAMPATH (Hamiltonian path)}  problem given  in \cite{sipser} to explain this \textit{NDTM} in the proof:

\textit{
The following is a nondeterministic Turing machine (NDTM) that decides the HAMPATH problem in nondeterministic polynomial time. Recall that in Definition 7.9 we defined the time of a nondeterministic machine to be the time used by the longest computation branch.\\ \\
N = " On input $<G, s, t>$, where G is a directed graph with nodes s and t:
 \begin{enumerate}
\item Write a list of m numbers, $p_1\dots p_m$, where m is the number of nodes in G. Each number in the list is nondeterministically selected to be between 1 and m.
\item Check for repetitions in the list. If any are found, reject.
\item Check whether $s=p_1$ and $t=p_m$. If either fail, reject.
\item For each i between 1 and m-1, check whether  $(p_i, p_{i+1})$ is an edge of G. If any are not, reject. Otherwise, all tests have been passed, so accept. "
 \end{enumerate}  
To analyze this algorithm and verify that is runs in nondeterministic polynomial time, we examine each of its stages. In stage 1, the nondeterministic selection clearly runs in polynomial time. In stage 2 and 3, each part is a simple check, so together they run in polynomial time. Finally, stage 4 also clearly runs in polynomial time. Thus this algorithm runs in nondeterministic polynomial time.
 }\\
 
When $p_1\dots p_m$ is checked to be a Hamiltonian path, the corresponding \textit{NDTM} $N$  \textit{accepts} the instance $<G, s, t>$, and determines that  the instance $<G, s, t>$ has a solution. But when $p_1\dots p_m$ is checked not to be a Hamiltonian path,  \textit{NDTM} $N$ can neither determine that the instance $<G, s, t>$ has no solution, nor determine that the instance $<G, s, t>$ has a solution, because $p_1\dots p_m$ is just a certificate.  In this case,   the decision for accepting $<G, s, t>$ is \textit{nondeterministic}. In other words, the \textit{verification} is not  consistent  with the \textit{decision}. 

Therefore,  the \textit{verification} cannot be used to define  \textit{NP},  and the equivalence of the two definitions of NP does not hold!

On the other hand,  if people insist the equivalence of the two definitions of \textit{NP}, then it means  the logic error of disguised displacement would be allowed to exist. Consequently the \textit{verification} of  \textit{TM}   would be confused up with the  \textit{transcendent  judgement} of  \textit{Oracle},  finally replace the   \textit{nondeterministic decision}  about \textit{NP},  while the   \textit{nondeterministic decision}  is just the essence of \textit{NP}. All this is what happens actually  in the theory of complexity of algorithms.

\section{Conclusion}

In this paper, we  revealed   the disguised displacement of concept   \textit{NDTM}     in the definition of  \textit{NP}, which causes ambiguities in understanding  \textit{NP} and finally  great  difficulties in solving the  \textit{P versus NP} problem.

Since  \textit{NP} is originally related with $Oracle$ that comes from Turing's work about  \textit{Computability}, it seems quite necessary to trace back to  Turing's work and  clarify further   the issue about \textit{NP} \cite{scott}. 


\end{document}